\documentclass[12pt]{article}
\usepackage[dvipsnames]{xcolor}
\usepackage[colorlinks=true,
linkcolor=Blue, 
citecolor=Blue,
filecolor=Blue,
urlcolor=Blue,
linktoc=page, 
pdfstartview=FitV,
bookmarksopen=true]{hyperref}
\usepackage{amsmath, amssymb, amsfonts, graphicx, cite,tikz, spectralsequences, braket}
\usepackage[a4paper, left=2.5cm, right=2.5cm, top= 3cm, bottom=4cm]{geometry}

\usetikzlibrary{arrows}

\DeclareSseqGroup \tower {} {\class(0,0)\DoUntilOutOfBounds {\class(\lastx,\lasty+1)\structline}}

\def\sqtwoR (#1,#2,#3){\draw[#3] (#1,#2) .. controls (#1+1,#2+1) .. (#1,#2+2);}
\def \sqtwoCR (#1,#2,#3){\draw[#3] (#1,#2) .. controls (#1+1,#2+.5) and (#1+1.5,#2+2) .. (#1+2,#2+2);}
\def \sqone (#1,#2,#3){\draw[#3] (#1,#2) -- (#1,#2+1);}

\def\a{\alpha}
\def\b{\beta}
\def\G{\Gamma}

\def\s{\sigma}
\def\S{\Sigma}

\def\cT{\mathcal{T}}
\def\cG{\mathcal{G}}
\def\cD{\mathcal{D}}
\def\cE{\mathcal{E}}
\def\cL{\mathcal{L}}
\def\cA{\mathcal{A}}
\def\cO{\mathcal{O}}
\def\cM{\mathcal{M}} 
\def\bP{\mathbb{P}}
\def\bZ{\mathbb{Z}}
\def\bR{\mathbb{R}}
\def\bC{\mathbb{C}}
\DeclareMathOperator{\Hom}{Hom}
\DeclareMathOperator{\Det}{Det}
\DeclareMathOperator{\Ind}{Ind}
\DeclareMathOperator{\Spec}{Spec}
\DeclareMathOperator{\sign}{sign}

\numberwithin{equation}{section}

\begin{document}
\begin{titlepage}

\vspace*{0.8cm} 
\begin{center}
{\LARGE 
Dai-Freed anomalies and level matching in \\heterotic asymmetric orbifolds
}\\
\vspace*{1.5cm}
Peng Cheng$^{1}$ and H\'ector Parra De Freitas$^2$  \\

\vspace*{1.0cm} 
$^1$ {\it Johannes Gutenberg-Universit\"at, Staudinger Weg 7, 55128 Mainz, Germany}\\

$^2$ {\it  Department of Physics, Harvard University, Cambridge, MA 02138, USA}\\

{pcheng@uni-mainz.de,  hparradefreitas@fas.harvard.edu}

\vspace{1cm}

\small{\bf Abstract} \\[3mm]\end{center}
We study asymmetric orbifolds of the $E_8\times E_8$ heterotic string from the perspective of worldsheet Dai-Freed anomalies. Focusing on cyclic symmetries $G = \bZ_m$ that act chirally on the fermions and symmetrically on the bosons, we compute the corresponding spin-bordism invariants and derive the conditions for the vanishing of global anomalies from this perspective. In the fermionic description, these conditions are exactly the familiar level-matching constraints, together with the additional mod-2 conditions that appear for even $m$. We then discuss the same conditions from the transformation properties of higher-genus fermion partition functions and explain how the anomaly is matched under bosonization for a large class of inner automorphisms of the $E_8\times E_8$ lattice theory. This gives an interpretation of the standard consistency conditions for asymmetric heterotic orbifolds from the Dai-Freed anomaly perspective.
\end{titlepage}
\newpage
\tableofcontents	

\section{Introduction}
\label{sec:intro}
Orbifolds furnish one of the basic mechanisms for constructing perturbative string vacua \cite{Dixon:1985jw,Dixon:1986jc}. Given a two-dimensional conformal field theory $\cT$ with finite symmetry group $G$, one can ask two standard questions: when is gauging $G$ consistent, and when it is consistent, how many inequivalent gaugings are possible? In the traditional orbifold literature \cite{Narain:1986qm,Narain:1990mw,Vafa:1986wx,Freed:1987qk}, the first question is controlled by level matching, while the second is controlled by discrete torsion.

Recent work on generalized global anomalies recasts these questions in bordism-theoretic terms \cite{Garcia-Etxebarria:2018ajm,Lee:2020ojw,Kaidi:2019tyf,BoyleSmith:2023xkd}. For a spin two-dimensional theory, the anomaly of $G$ has a perturbative part measured by $\Hom(\Omega_4^{\text{Spin,free}}(BG),\bZ)$ and a global part measured by 
$\Hom(\Omega_3^{\text{Spin,tor}}(BG),\bZ)$. From this perspective, gauging $G$ is allowed precisely when the corresponding worldsheet anomaly vanishes, while different gaugings are parametrized by a generalized version of discrete torsion related to bordism groups as well.

In this paper we apply this viewpoint to asymmetric orbifolds of the 
$E_8\times E_8$ heterotic string. We restrict to cyclic symmetries $G = \bZ_m$ that act chirally on the worldsheet fermions and symmetrically on the bosons, and we analyze the anomaly of the pre-GSO theory together with the GSO symmetries. Our main result is that the bordism criterion for anomaly cancellation reproduces exactly the standard level-matching condition in the fermionic description. For even $m$, the same analysis also yields the additional mod-2 constraints associated with mixed anomalies involving the GSO projections. We then recover the same conditions from higher-genus fermion partition functions and explain how they are matched in the bosonic $E_8$ lattice description for inner automorphisms preserved by fermionization.

This paper is organized as follows. In Section \ref{sec:fermanom} we derive the anomaly-free conditions for $G$ from the bordism analysis of the fermionic heterotic $E_8\times E_8$ worldsheet theory. In Section \ref{sec:theta} we recover the same conditions directly from the relevant part of the partition function of chiral fermions on a general Riemann surface $\Sigma_g$, using the family index theorem. In Section \ref{sec:bosonization} we compare the bosonic and fermionic descriptions and show how the anomaly of a general $\bZ_m$ symmetry is matched whenever that symmetry survives bosonization/fermionization. In Section \ref{sec:conclusion} we summarize the results and discuss several directions for future work. 

\section{Anomalies from Fermions}
\label{sec:fermanom}
We will work with the two-dimensional worldsheet theory $\cT_\text{HE}$ of the $E_8\times E_8$ heterotic string. In the fermionic description, the worldsheet fields are:
\begin{enumerate}
	\item Eight non-chiral physical bosons $X^i$, with $i = 1,...,8$.
	\item Eight right-moving Majorana--Weyl (MW) fermions $\psi_R^i$, grouped into four Weyl fermions $\xi_R^I$ with $I = 1,2,3,4$.
	\item Two sets of eight left-moving Weyl fermions $\lambda_L^A$, $\lambda_L^{A'}$, with $A,A' = 1,...,8$.
\end{enumerate}
There is moreover the usual GSO projection, equivalent to gauging the symmetry group 
\begin{equation}
	G_\text{GSO} = \bZ_2^{F_L}\times \bZ_2^{F_L'}\times \bZ_2^{F_R}
\end{equation}
generated by the fermion number operators
\begin{equation}\label{eq:gso}
	(-1)^{F_L}: ~ \lambda_L^A \mapsto -\lambda_L^A\,, ~~~~~~~ (-1)^{F_L'}: ~ \lambda_L^{A'} \mapsto -\lambda_L^{A'}\,, ~~~~~~~ (-1)^{F_R}: ~ \xi_R^I \mapsto -\xi_R^I\,.
\end{equation} 

Here we restrict attention to anomalies arising purely from chiral fermions. Accordingly, we consider only orbifold actions for which the action of $G$ on the bosons $X^i$ is non-chiral (symmetric). A nontrivial action of $G$ on the $\xi_R^I$ is understood to be paired with a corresponding action on the $X^i$, as required by worldsheet supersymmetry; in that case the $X^i$ are taken to be compact. For simplicity, we further restrict $G$ to be cyclic:
\begin{equation}
	G = \bZ_m\,, ~~~~~ 
\end{equation}
with arbitrary $m\geq 2$, and with generator $g$ acting on the worldsheet fermions as
\begin{equation}\label{eq:g-action-fermions}
    g: ~~~~~ \lambda_L^A \to e^{2\pi i q_A/m}\lambda_L^A\,, ~~~~~ \lambda_L^{A'} \to e^{2\pi i q_{A'}/m}\lambda_L^{A'}\,, ~~~~~ \xi_R^I \to e^{2\pi i q_I/m}\xi_R^I\,. ~~~~~ 
\end{equation}
Note that this choice excludes symmetries which are not connected to the identity. We will come back to this point in Section \ref{subsec:outer-automorphisms}.

In our analysis we require that $G$ is a symmetry both of $\cT_\text{HE}$ and of the pre-GSO theory $\cT$, and we study its anomaly in $\cT$. Equivalently, we require that $G$ has no mixed anomaly with any of the GSO involutions. In practice, we therefore compute the anomaly of the group $G'$ obtained by combining $G$ with $G_\text{GSO}$. From \eqref{eq:g-action-fermions} and \eqref{eq:gso}, $G$ commutes with $G_\text{GSO}$, so $G'$ is abelian:
\begin{equation}
    G' = G \times G_\text{GSO} = \bZ_m \times \bZ_2^{F_L}\times \bZ_2^{F_L'}\times \bZ_2^{F_R}\,.
\end{equation}
Moreover, since there are no mixed anomalies between $\bZ_2^{F_L}$, $\bZ_2^{F_L'}$ and $\bZ_2^{F_R}$, we can look separately at the anomalies of the groups
\begin{equation}
    \bZ_m \times \bZ_2^F\,,
\end{equation}
where $\bZ_2^F$ is shorthand for any of the GSO involutions.

\subsection{General considerations}
In this subsection we isolate the pure $\bZ_m$ anomaly in the fermionic description. We start by studying the anomaly $\cA_{\cT}\in\mathcal{A}$ of the symmetry group $G=\bZ_m\subset G'$. The group $\cA$ of potential anomalies fits into a short exact sequence:
\begin{equation}
	\label{eq:anom-2d-zm}
	0 \to \Hom_{\bZ}(\Omega_{3d}^{Spin,tor}(B\bZ_m),U(1)) \to \cA \xrightarrow{\pi} \Hom(\Omega_{4d}^{Spin,free}(B\bZ_m),\bZ) \to 0\,.
\end{equation}
In particular, the perturbative gravitational anomaly is encoded in the projection
\begin{equation}
	\label{eq:anom-grav-T}
	\pi(\cA_\cT) = \frac{16-4}{24}p_1 \in  \Hom(\Omega_{4d}^{Spin,free}(B\bZ_m),\bZ)\,,
\end{equation}
and is sensitive only to the difference in number between left-moving and right-moving fields. Since we are interested in the non-perturbative $\bZ_m$ anomaly, we eliminate the perturbative part \eqref{eq:anom-grav-T} by introducing four left-moving Weyl fermions $\xi_L^I$ and sixteen right-moving Weyl fermions $\lambda_R^A, \lambda_R^{A'}$, all neutral under $\bZ_m$. The $\bZ_m$ anomaly is then
\begin{equation}
	\label{eq:elim-grav-anom}
	\cA_\cT - \pi(\cA_\cT) \in \Hom_{\bZ}(\Omega_{3d}^{Spin,tor}(B\bZ_m),U(1))\,.
\end{equation}

It is well known that \eqref{eq:elim-grav-anom} is given by the invertible field theory coming from massive $3d$ Dirac fermions with $\bZ_m$ charges. For a $2d$ Weyl fermion $\psi_{\pm,q}$ with $\bZ_m$ charge $q$ we have the map 
\begin{equation}
	\label{eq:ift-map}
	\cA_{\psi_{\pm,q}} -\pi(\cA_{\psi_{\pm,q}}) =  \pm(\eta_q - \eta_{0})\,.
\end{equation}
$\eta_q$ is the eta invariant of the $3d$ Dirac operator $\cD_{3d,q}$ coupled to the $\bZ_m$-bundle, coming from a massive $3d$ Dirac fermion with $\bZ_m$ charge $q$, and is defined as 
\begin{equation}
	\label{eq:eta-def}
	\eta_q = \sum_{\lambda \in \Spec(\cD_{3d,q})}\frac{\sign(\lambda)}{2}, ~~~~~~~~ \sign(0) = 1.
\end{equation}
Although by itself $\eta_q$ is a geometric invariant, i.e. it depends on local geometric data \cite{atiyah1975spectral}, subtracting $\eta_0$ cancels out this dependence thus giving a bordism invariant valued in $\Hom_{\bZ}(\Omega_{3d}^{Spin,tor}(B\bZ_m),U(1))$, see\cite{atiyah1975spectral2}. 

The map \eqref{eq:elim-grav-anom} is indeed compatible with the general description of anomalies \eqref{eq:anom-2d-zm}. Let $X_3$ be a three-dimensional compact spin manifold with $\mathbb{Z}_m$-structure. In the case that $X_3$ is null-bordant, the Atiyah-Patodi-Singer (APS) theorem gives
\begin{equation}
	\eta_q = \int_{X_4} \hat A(X) = -\frac{1}{24}\int_{X_4}p_1 \mod \bZ\,, 
\end{equation}  
with $\partial X_4 = X_3$ for all $q$. The RHS is independent of the $\bZ_m$-charge, and gives precisely the gravitational anomaly of a 2d Weyl fermion. In the case that $X_3$ is not null-bordant, the combination $\pm (\eta_q-\eta_0)$ gives instead the pure $\bZ_m$-anomaly as discussed above. For more details and examples about the bordism description of anomalies, see \cite{Garcia-Etxebarria:2018ajm,Lee:2020ojw,Cheng:2025ikd,Kaidi:2019tyf,Tachikawa:2018njr,Dierigl:2026sok, Larotonda:2026hxy}.

It follows that, for our theory $\cT$, the $\bZ_m$ anomaly is given by
\begin{equation}\label{eq:zm-anom-T}
	\widetilde\cA_\cT = \sum_{A = 1}^8 \tilde\eta_{q_A} + \sum_{A' = 1}^8 \tilde\eta_{q_{A'}} -  \sum_{I = 1}^4 \tilde\eta_{q_I}\,,
\end{equation} 
with  $\widetilde{\cA}_\cT \equiv \cA_\cT - \pi(\cA_\cT)$ and  $\tilde \eta_q \equiv \eta_q - \eta_0$. 

As discussed above, we are interested more generally in the anomaly for $\bZ_m \times \bZ_2^F$, namely
\begin{equation}
	\label{eq:zm-z2-anom-T}
	\widetilde\cA_\cT \in \Hom_{\bZ}(\Omega_{3d}^{Spin,tor}(B\bZ_m\times B\bZ^F_2),U(1))\,.
\end{equation}
In the following we will determine the conditions under which $\widetilde{\cA}_\cT$ vanishes. 
This requires (1) determining the generators of $\Omega_{3d}^{Spin,tor}(B\bZ_m\times B\bZ^F_2)$ and (2) evaluating \eqref{eq:zm-anom-T} on each of them. Our strategy will be to compute the anomaly-free condition for the symmetries $\bZ_m$ and $\bZ_2^F$ separately and then determine the vanishing condition for the mixed anomaly, if there is any. 

\subsection{$\bZ_2^F$ anomaly}
\label{subsec:gso-anom}
We first recall the well-known $\bZ_2^F$ anomaly constraint. In this case, the anomaly-free condition is already well studied \cite{Kaidi:2019tyf}. The relevant bordism group is
\begin{equation}
	\Omega_{3d}^{Spin,tor}(B\bZ_2) = \bZ_8\,,
\end{equation}
generated by $\bR \bP^3$ with a flat $\bZ_2$ background. The $\bZ_2$ anomaly on the generator is
\begin{equation}
	\eta_1 - \eta_0 = \frac14\,.
\end{equation}
Letting $\bZ_2^F$ act as -1 on $n$ Weyl fermions, the anomaly-free condition reads
\begin{equation}\label{eq:mod4}
	n(\eta_1-\eta_0) \in \mathbb{Z} ~~~\Rightarrow~~~ n \in 4\bZ\,.
\end{equation}
The 2d theory must then have a multiple of four Weyl fermions charged under the symmetry $\bZ_2^F$ for it to be anomaly-free. This condition is clearly satisfied in $\cT$ for $\bZ_2^{F_L}$, $\bZ_2^{F_L'}$ and $\bZ_2^{F_R}$ since $n = 8,8,4$, respectively.

Note that $\eta_1-\eta_0 = 1/4$ is \textit{twice} the generator of $\Hom_{\bZ}(\Omega_{3d}^{Spin,tor}(B\bZ_2),U(1)) = \bZ_8$. This is so because the eta invariants count the anomaly contribution from a Weyl fermion, which is reducible into two MW fermions as a representation of $\bZ_2$. The mod 4 condition \eqref{eq:mod4} is thus equivalent to the more familiar mod 8 condition for MW fermions. For $\bZ_{m > 2}$, however, the group action is always represented by complex phases under which Weyl fermions are irreducible. For this reason we do not use MW fermions in our analysis. 

\subsection{$\bZ_{m \geq 2}$ anomaly}
We next compute the $\bZ_m$ anomaly by evaluating eta invariants on generators of the relevant bordism group. Let us factorize $m = 2^k p$ with $p$ odd. 
The form of the bordism group $\Omega_{3d}^{Spin,tor}(B\bZ_m)$ falls then into three distinct classes depending on the value of $k$, as summarized in the following table (see e.g. appendix C of \cite{Cheng:2025ikd})
\begin{align}
	\begin{tabular}{|c|c|c|}\hline
		$m$&$\Omega_{3d}^{Spin,tor}(B\bZ_m)$&generators\\\hline\hline
		$p$&$\bZ_p$&$L_p^3(1)$\\
		\hline
		$2p$&$\bZ_8 \oplus \bZ_p$&$L_2^3(1),~L_p^3(1)$\\
		\hline
		$2^{k \geq 2}p$&$\bZ_{2m} \oplus \bZ_2	$&$L_m^3(1),~T^3(1)$\\
		\hline
	\end{tabular}
\end{align}
Here $L_r^3(s)$ is the lens space $S^3/\bZ_r$ with $\bZ_r$ holonomy $e^{2\pi i s/r}$, and $T^3(1) = S^1 \times T^2$ where $S^1$ carries a $\bZ_m$ holonomy $e^{2\pi i/m}$ and $T^2$ has periodic Spin structure. 

To compute the anomalies we evaluate the reduced eta invariant $\tilde \eta_q$ on the generators listed above. For the lens spaces we have the formula (see e.g. \cite{Monnier:2018nfs,Dierigl:2025rfn})
\begin{equation}
	\tilde \eta_q(L_p^3(s))  = \frac{sq(sq-p)}{2p}\,.
\end{equation}
For $T^3(1)$, we use the product formula
\begin{equation}
\label{eq:eta-prod}
	\eta_q^{3d}(S^1\times T^2) = \eta_q^{1d}(S^1)\cdot \Ind(T^2) = 0 ~~~~~ \Rightarrow ~~~~~ \tilde\eta_q(T^3(1)) = 0\,,
\end{equation}
where we use $\Ind(T^2) = 0$. We thus have 
\begin{equation}
	\widetilde{\cA}_\cT(L_r^3(s)) = \frac{1}{2r}\left(\sum_{A = 1}^8 sq_A(sq_A-r) + \sum_{A' = 1}^8 sq_{A'}(sq_{A'}-r) - \sum_{I = 1}^4 sq_I(sq_I-r)\right) \mod 1
\end{equation}
and 
\begin{equation}
	\widetilde{\cA}_\cT(T^3(1)) = 0 \mod 1\,.
\end{equation}
All of the anomaly cancellation conditions are then of the form
\begin{equation}\label{eq:anom-gen}
	\sum_{A = 1}^8 sq_A(sq_A-r) + \sum_{A' = 1}^8 sq_{A'}(sq_{A'}-r) - \sum_{I = 1}^4 sq_I(sq_I-r) = 0 \mod 2r\,.
\end{equation}

For the case $m = p$ odd, note that the product $sq(sq-p)$ is necessarily even, and we have the implication
\begin{equation}
	sq(sq-p) = 0 \mod p ~~~~~\Rightarrow~~~~~ sq(sq-p) = 0 \mod 2p\,.
\end{equation}
Taking $s = 1$ and using $q \in \mathbb{Z}$,  eq. \eqref{eq:anom-gen} then reduces to
\begin{equation}
		Q^2 \equiv \sum_{A = 1}^8 q_A^2 + \sum_{A' = 1}^8 q_{A'}^2 - \sum_{I = 1}^4 q_I^2 = 0 \mod p\,.
\end{equation}
For $m = 2p$, the two conditions can be reduced respectively to $Q^2 = 0 \mod p$ and $Q^2 = 0 \mod 4$. Equivalently, $2Q^2$ must be divisible by $p$ and 8 simultaneously, and since $p$ is odd, it must be divisible by $8p$, hence $Q^2 = 0  \mod 4p$. Similarly for $m = 2^{k\geq2}p$ we find a mod $2m$ condition, and so the overall anomaly cancellation condition for the $\bZ_m$ symmetry reads
\begin{equation}
\label{eq:zm-anom}
	\sum_{A = 1}^8 q_A^2 + \sum_{A' = 1}^8 q_{A'}^2 - \sum_{I = 1}^4 q_I^2  = 
	\begin{cases}
		0 \mod m & m~\text{odd}\\
		0 \mod 2m & m~\text{even}\\
	\end{cases}\,.
\end{equation} 

\subsection{$\bZ_m\times \bZ_2^F$ anomaly}
Finally, we analyze the mixed anomaly between $\bZ_m$ and $\bZ_2^F$. The full anomaly for $G'=\bZ_m\times \bZ_2^F$ splits into the anomalies for $\bZ_m$ and $\bZ_2^F$ computed above, together with their mixed anomaly. To see this in the language of bordism, we note first that there are two sequence of maps
\begin{align}
	B\bZ_n &\overset{i}{\longrightarrow} B\bZ_n \times B\bZ_2^F \overset{\pi_1}{\longrightarrow} B\bZ_n\,,\label{eq:ret1}\\
	B\bZ_2^F &\overset{i}{\longrightarrow} B\bZ_n \times B\bZ_2^F \overset{\pi_2}{\longrightarrow} B\bZ_2^F\,,\label{eq:ret2}
\end{align}
with $ \pi_1\circ i$ and $ \pi_2 \circ i$ identity maps. From standard results in generalized homology theory \cite{davis2001lecture} we have that for $X$ a retract of $Y$, 
\begin{equation}\label{eq:gen-hom-retract}
\Omega_{3d}^{Spin,tor}(Y) = \Omega_{3d}^{Spin,tor}(X)\oplus G_X\,,	
\end{equation}
with $G_X$ a group to be determined. Applying \eqref{eq:gen-hom-retract} to \eqref{eq:ret1} and \eqref{eq:ret2} it follows that
\begin{equation}\label{eq:bordism-K}
	\Omega_{3d}^{Spin,tor}(B\bZ_m\times B\bZ_2^F)) = \Omega_{3d}^{Spin,tor}(B\bZ_m) \oplus \Omega_{3d}^{Spin,tor}(B\bZ_2^F) \oplus K\,.
\end{equation}
We leave the explicit computation of $K$ to appendix \ref{app:bordism-computation}. In summary, we find 
\begin{align}
	\begin{tabular}{|c|c|c|}\hline
		$m$&$K$&generators\\\hline\hline
		$p$&$0$&-\\
		\hline
		$2p$&$\bZ_4$&$ X_{3d}$\\
		\hline
		$2^{k \geq 2}p$&$\bZ_2 \oplus \bZ_2	$&$ Y_{1,3d}, Y_{2,3d}$\\
		\hline
	\end{tabular}
\end{align}
The generators $X_{3d}$ and $Y_{i,3d}$ are respectively a circle bundle over the $\text{Pin}^-$ manifold $\bR\bP^2$ and $S^1\times T^2$, both with suitable $\bZ_2\times \bZ_2$ bundle structures.\footnote{The group $K = \bZ_4$ is associated to the quadratic enhancement of the intersection form on a 2d $\text{Pin}^-$ manifold, see e.g. \cite{Guo:2018vij, Kaidi:2019tyf}.} 

Since $K = 0$ for $m = p$ and eta invariants vanish on $S^1\times T^2$ (cf. eq. \eqref{eq:eta-prod}), the potential new constraint from $G_X$ only arise for $m = 2p$. In this case we can write
\begin{equation}
    G' = \bZ_{2p} \times \bZ_2^F = \bZ_p \times \bZ_2 \times \bZ_2^F\,,
\end{equation}
and from our discussion above, $\bZ_p$ has no mixed anomaly with either $\bZ_2$ or $\bZ_2^F$, hence there can only be a potential mixed anomaly in $\bZ_2 \times \bZ_2^F \subset G'$. Since $K = \bZ_4$, and $\bZ_2\times \bZ_2^F$ charges can be carried by MW fermions, we see that for 2d Weyl fermions $K$ can give at most a mod 2 condition by evaluating \eqref{eq:anom-gen} on $X_{3d}$. 

To complete our analysis and clarify what is the above mod 2 condition, we now consider further possible constraints coming from having three sets of Weyl fermions charged under different GSO involutions. In the above analysis of the $\bZ_m$ symmetry, the Spin structures of the lens spaces $L_r^3(s)$ are taken to be the same when evaluating \eqref{eq:anom-gen} for each 2d Weyl fermion. When combined with $G_\text{GSO}$, however, the Spin structures associated to the three sets of Weyl fermions are independent from each other, as well as the associated right moving Weyl fermions introduced to cancel the perturbative anomalies. We must evaluate \eqref{eq:anom-gen} on each possible choice.

For $m = p$, we have $H^{1}(L_{p}^{3}(s),\bZ_2) = 0$, hence there is only one overall choice of Spin structures, thus no new constraints. For both $m = 2p$ and $m = 2^{k \geq 2}p$, however, we have non-trivial choices coming from $H^1(L_2^3(1),\bZ_2) = H^1(L^3_{m= 2^{k \geq 2}p}(1),\bZ_2) = \bZ_2$. If, for example, we choose different Spin structures for $(\lambda_L^A$, $\lambda_L^{A'})$ and $\xi_R^I$, condition \eqref{eq:anom-gen} is modified to
\begin{equation}
\sum_{A = 1}^8 sq_A(sq_A-m) + \sum_{A' = 1}^8 sq_{A'}(sq_{A'}-m) - \sum_{I = 1}^4 (sq_I+\frac{m}{2})(sq_I-\frac{m}{2}) = 0 \mod 2m\,.
\end{equation}
Combining with the original \eqref{eq:anom-gen}, we obtain
\begin{equation}
\label{eq:new-anom-m2p-1}
 \forall s , sm\sum_{I=1}^4 q_I = 0 \mod 2m  ~~~~~\Rightarrow~~~~~ \sum_{I=1}^4q_I = 0 \mod 2,
\end{equation}
and including the rest of Spin structure choices we get three new mod 2 conditions
\begin{equation}
\label{eq:new-anom-m2p-2}
\sum_{A=1}^8 q_{A} =\sum_{A'=1}^8 q_{A'}= \sum_{I=1}^4q_I = 0 \mod 2.
\end{equation}
In the case $m = 2p$, these are precisely the mod 2 conditions coming from $K$, one for each $\bZ_2 \times \bZ_2^F$ group associated to a set of Weyl fermions. 

In summary, we have found that $G' = \bZ_m \times G_\text{GSO}$ in $\cT$ and the corresponding $G = \bZ_m$ in $\cT_\text{HE} = \cT/G_\text{GSO}$ are free of anomalies if
\begin{equation}
\label{eq:full-anomaly}
\begin{split}
& \sum_{A = 1}^8 q_A^2 + \sum_{A' = 1}^8 q_{A'}^2 - \sum_{I = 1}^4 q_I^2  = 0 \mod m \quad \text{$m$ odd},\\&
\begin{cases}
\sum_{A = 1}^8 q_A^2 + \sum_{A' = 1}^8 q_{A'}^2 - \sum_{I = 1}^4 q_I^2  = 0 \mod 2m  \\
\sum_{A=1}^8 q_{A} =\sum_{A'=1}^8 q_{A'}= \sum_{I=1}^4q_I = 0 \mod 2.
\end{cases},\quad \text{$m$ even.}
\end{split}
\end{equation}
These are precisely the conditions found in \cite{Vafa:1986wx} for obtaining a level-matched spectrum in the orbifold theory $\cT_\text{HE}/G$. We have therefore shown that the same conditions also ensure the vanishing of the nonperturbative gravitational anomaly. The discussion extends in a straightforward way to other finite abelian groups and to the $Spin(32)/\bZ_2$ heterotic string.

An immediate consequence of \eqref{eq:full-anomaly} is that, for a theory consisting of $n$ left-moving Weyl fermions with symmetry group $G=\bZ_m\times \bZ_2^F$, the 't~Hooft anomaly-free condition is
\begin{equation}\label{eq:anom-free-general-fermions}
\begin{cases}
n = 0 \mod 4\\
\sum_{i=1}^{n}q_i^2 = 0 \mod m \quad \text{$m$ odd}\\
\sum_{i=1}^{n}q_i^2 = 0 \mod 2m \quad \text{$m$ even}
\end{cases}
\end{equation}

\paragraph{Remark} Consider a theory of $n$ left-moving Weyl fermions with symmetry group $G=\bZ_m$. Then the 't~Hooft anomaly-free condition becomes
\begin{equation}
\begin{cases}
 \sum_{i=1}^{n}q_i^2 = 0 \mod m \quad \text{$m$ odd}\\
\sum_{i=1}^{n}q_i^2 = 0 \mod 2m \quad \text{$m$ even}.
\end{cases}
\end{equation}

The condition $n=0 \mod 4$ can now be dropped. However, when $m$ is even, $\bZ_m$ contains a $\bZ_2$ subgroup, so one expects a mod-$4$ constraint from that subgroup. This condition is already encoded in the $\bZ_m$ anomaly-free condition. Indeed, the $i$th fermion is charged under this $\bZ_2$ subgroup precisely when $q_i\neq 0 \mod 2$, and we have:
\begin{equation}
\begin{split}
 \sum_{i=1}^{n}q_i^2 = 0 \mod 2m \to 
\sum_{i=1}^{n}q_i^2 = 0 \mod 4 \to 
\sum_{i:q_i \neq 0 \mod 2} q_i^2 = 0\mod 4.
\end{split}
\end{equation}
Thus the mod-$4$ condition from the $\bZ_2$ subgroup is already encoded in the $\bZ_m$ anomaly-free condition. 
\section{Theta function and fermion partition function on $\S_g$}
\label{sec:theta}

In this section we summarize a first-principles derivation of the chiral-fermion partition function from a global point of view, and then study the coupling to background $\bZ_m$ fields in order to recover the anomaly-free conditions derived in the previous section. Although the results are well known \cite{Alvarez-Gaume:1986rcs}, the geometric viewpoint used here may be less familiar and helps clarify the origin of the anomaly constraints. 

\subsection{Partition function of chiral fermions on general Riemann surface}
\label{subsec:fermion-partition-general}

A chiral fermion $\psi_{+}$ is valued in the sections of the line bundle $K^{1/2}\otimes \cL$, with $K^{1/2}$ the spin bundle and $\cL$ the flavor $U(1)$ symmetry of the complex chiral fermions, and the Dirac operator is given by
\begin{equation}
    \label{eq:dirac-2d}
    i\bar{\partial}: \G(K^{1/2}\otimes \cL) \to \G(K^{-1/2}\otimes \cL)\,.
\end{equation}
For a Riemann surface $\Sigma_g$ of genus $g$, the action with general flat $U(1)$ flavor background is of the form
\begin{equation}
    \label{eq:fermion-action}
    S_{ferm} = \int_{\Sigma_g}\sqrt{h}d^2 \sigma_g \bar{\psi}_{+}i(\bar{\partial}_A) \psi_{+}\,, ~~~~~\bar{\partial}_A \equiv \bar{\partial} + A_{\bar{z}}\,, 
\end{equation}
with $A_{\bar{z}}$ the antiholomorphic component of the $U(1)$ background field. The partition function at $g$ loops is then 
\begin{equation}
    \label{eq:fermion-partition}
    Z_{ferm} = \int{D\bar{\psi}D\psi} e^{-S_{ferm}} = \Det(i\bar{\partial}_A)\,,
\end{equation}
where $\Det(i\bar{\partial}_A)$ can be viewed as a section of a line bundle $\cG$ on the moduli space $\cM_{flat}$ of flat line bundles on $\Sigma_{g}$. 

\subsubsection{Moduli space $\cM_{flat}$ as Jacobian of $\Sigma_g$}
The moduli space $\cM_{flat}$ is the Jacobian $J(\Sigma_g)$ of $\Sigma_g$. We first describe its topology and then its structure as an abelian variety in terms of period matrices.  The first homology group is $H_1(\Sigma_g) = \bZ^{2g}$,  generated by cycles $A_{i=1,..,g}$ and $B_{i=1,..,g}$ with intersections
\begin{equation}
    \label{eq:ab-cycle-intersection}
    A_{i}\cap B_{j} = - B_{i}\cap A_{j} = \delta_{ij}, \quad A_{i}\cap A_{j} = B_{i}\cap B_{j} = 0\,.
\end{equation}
We denote the Poincaré duals of these cycles as $x_{i},y_{i} \in H^{1}(\Sigma_g;\bZ)$, i.e.
\begin{equation}
    \label{eq:ab-cycle-duals}
    \int_{A_i}x_j = \int_{B_i} y_j = \delta_{ij}, \quad \int_{A_i}y_j = \int_{B_i} x_j = 0. 
\end{equation}
The flat line bundle over $\S_g$ can then be characterized by its holonomies along $A_i$ and $B_i$, and the corresponding flat $U(1)$ connection is given by
\begin{equation}
    \label{eq:flat-u1-connection}
    A = 2\pi \a^{i}x_i - 2\pi \b^{i}y_i, \quad \a_i,\b_i \in \bR/\bZ\,.
\end{equation}
In particular, the chiral fermion coupled to the $U(1)$ flat line bundle satisfies the conditions
\begin{equation}
    \psi(\sigma+A_i) = e^{2\pi i\a_i}\psi({\sigma}),   ~~~~~~~~~~\psi(\sigma+B_i) = e^{-2\pi i\b_i}\psi({\sigma})\,,
\end{equation}
hence, topologically, we have
\begin{equation}
    \label{eq:mflat-topology}
    \cM_{flat} = \bR^{2g}/\bZ^{2g}\,.
\end{equation}

$\cM_{flat}$ is also an abelian variety $\bC^g/\bZ^g \oplus \Omega\bZ^{g}$, where $\Omega$ is a symmetric complex $g\times g$ matrix with positive imaginary part. This matrix is given by the periods
\begin{equation}
    \label{eq:period-matrix}
    \Omega_{ij}= \int_{B_i} w_{j}\,, 
\end{equation}
where $w_{1,..,g}$ is a basis of $H^{1}(\S_g;\cO)$ normalized by $A_{1,..,g}$ cycles:
\begin{equation}
    \label{eq:a-cycle-normalization}
    \int_{A_i}w_j = \delta_{ij}\,.
\end{equation}
Hence the moduli space $\cM_{flat}$ is an abelian variety $J(\S_g) = \bC^g/\bZ^g \oplus \Omega \bZ^g$.  Its complex structure is determined by the Riemann surface $\S_g$, and the holomorphic coordinates are given in terms of the antiholomorphic part $\bar{A}_i$ of the flat connection on $\S_g$ \eqref{eq:flat-u1-connection}:
\begin{equation}
    \label{eq:holomorphic-coordinates}
     A = 2\pi \a^{i}x_i - 2\pi \b^{i}y_i = A^{i}w_{i}+ \bar{A}^i \bar{w_i}\,.
\end{equation}

\subsubsection{Line bundle $\cG$ from family index theorem} We now describe the line bundle $\cG$ over $\cM_{flat}$ in which \eqref{eq:fermion-partition} takes values. Here we use the family index theorem \cite{bismut1986atiyah} to derive the first Chern class $c_1(\cG)\in H^2(\cM_{flat},\bZ)$ of $\cG$.

Take the overall space $\S_g \times \cM_{flat}$ and consider the $\cM_{flat}$ parametrization of the Dirac operator $i\bar{\partial}_A$ in \eqref{eq:fermion-action},
\begin{equation}
    \label{eq:mflat-dirac-family}
    (\a^i,\b^i) \in \cM_{flat} \to i\bar{\partial}_A = i(\bar{\partial} +\bar{A})\,,
\end{equation}
where $\bar{A}$ is given by \eqref{eq:holomorphic-coordinates}. The Family index theorem gives
\begin{equation}
    \label{eq:family-index-riemann}
    c_1(\cG) = \int_{\S_g}A(\S_g)ch(\cL)|_{4} =\sum_{i=1}^{g} d\a^i \wedge d\b^i \in H^{2}(\cM_{flat};\bZ)\,,
\end{equation}
where $\cL$ is the $U(1)$ bundle over $\S_g \times \cM_{flat}$ whose connection is given by \eqref{eq:flat-u1-connection}. As discussed before, $\cM_{flat}$ is an abelian variety and $\cG$ is a holomorphic line bundle\footnote{$\cG$ is an ample line bundle associated with Theta divisor of the Abelian variety.}, hence we can apply the Hirzebruch-Riemann-Roch theorem to compute the dimension of global sections, obtaining
\begin{equation}
    \label{eq:global-sections-G}
    H^0(\cM_{flat};\cG) = \int_{\cM_{flat}}td(\cM_{flat})e^{c_{1}(\cG)} = \frac{1}{g!}\int_{\cM_{flat}}(\sum_{i=1}^{g} d\a^i \wedge d\b^i)^g =1\,.
\end{equation}
Here we have used Kodaira's vanishing theorem and the fact that the orientation of $\cM_{flat}$ as an abelian variety is given by $\bigwedge_{i=1}^gd\a^i\wedge d\b^i$. 

\subsubsection{Fermion partition functions on $\S_{g\geq 1}$ and anomalies}
\label{subsec:fermion-partition-sg}
The global section of $\cG$ is unique up to an overall constant, and this unique section is given by theta functions \cite{beauville2013theta}. Since the fermion partition function on $\Sigma_g$, eq.~\eqref{eq:fermion-partition}, is valued in the global sections of $\cG$, it is uniquely determined by a theta function up to an overall factor. This overall factor is irrelevant for the dependence of \eqref{eq:fermion-partition} on the $U(1)$ flavor background, but it is important for the full partition function because it encodes the gravitational anomaly of the chiral fermion.

As the unique section of the degree $1$ line bundle over $\cM_{flat}$, up to an overall factor, the theta function $\vartheta(z|\Omega)$ is given by
 \begin{equation}
     \label{eq:theta-function}
     \vartheta(z|\Omega) = \sum_{n\in \bZ^g}\exp\{\pi i n^{T}\cdot \Omega \cdot n + 2\pi i n^{T}\cdot z\}, \quad \forall z \in \bC^g/\bZ^g\oplus \Omega\bZ_g\,.
 \end{equation}
 This function can be refined by characteristics. By applying the above procedure of deriving fermion partition functions to fermions with twisted boundary conditions
 \begin{equation}
 \label{eq:twisted-fermions}
    \psi(\s+A_i) = e^{ia_i}\psi(\s), \quad  \psi(\s+B_i) = e^{-ib_i}\psi(\s)\,,
 \end{equation}
 one obtains
\begin{equation}
    \label{eq:theta-characteristic}
    \vartheta\begin{bmatrix}
a \\[.1pt]
-b 
\end{bmatrix}(z|\Omega)  = \sum_{n\in \bZ^g}\exp\{\pi i (n+a)^{T}\cdot \Omega \cdot (n+a) + 2\pi i (n+a)^{T}\cdot (z-b)\}\,,
\end{equation}
with $z \in \bC^g/\bZ^g\oplus \Omega\bZ_g$ and $a,b \in \mathbb{R}^g/\mathbb{Z}^g$. 

There are two cases of interest for us. The first case comprises the $2^g$ choices $a_{1,..,g},b_{1,..,g} \in\{0,\frac{1}{2}\}$,  each of which defines a spin structure on $\S_g$ via \eqref{eq:twisted-fermions}, and is called a theta characteristic in algebraic geometry. In the second case the fermionic CFT has a $\bZ_m$ flavor symmetry, and the partition function with non-trivial flat $\bZ_m$ background on $\S_g$ is given by \eqref{eq:theta-characteristic} with $a_{1,..,g},b_{1,..,g} \in \frac{\bZ}{m} \mod \bZ$.  

 To summarize, the partition function of a single chiral fermion on $\S_{g\geq 1}$ with flat $U(1)$ flavor background specified by \eqref{eq:twisted-fermions} is given by
 \begin{equation}
 \label{eq:single-fermion-partition-sg}
 Z_{T_{n_F=1}}[\S_g,(a_i,b_i)] = c  \cdot \vartheta\begin{bmatrix}
a \\[.1pt]
-b 
\end{bmatrix}(z=0|\Omega)\,.  
 \end{equation}
 Here $c$ is independent of the $U(1)$ background, but depends on the complex structure of $\S_g$ and encodes the (perturbative) gravitational anomaly of the chiral fermion. For the purposes of this paper, we focus on the $U(1)$ background dependent part and ignore this overall factor.

\eqref{eq:single-fermion-partition-sg} can be generalized to the case $n_F > 1$ of multiple chiral fermions. Let us denote the corresponding theory by $T_{n_F}$, and the $\bZ_m$ charge of each fermion by $q_I$ with $I = 1,...,n_F$. Embedding the $\bZ_m$ holonomy into $U(1)$, the partition function of $T_{n_F}$ reads
\begin{equation} 
\label{eq:many-fermion-partition-sg}
Z_{{n_F}}[\S_g,(a_i,b_i)] =  c  ^{n_F} \cdot \prod_{I=1}^{T_{n_F}}\vartheta\begin{bmatrix}
q_{I}a \\[.1pt]
-q_{I}b 
\end{bmatrix}(z=0|\Omega)\,.
\end{equation}

This partition function can be used to study the $\bZ_m$ 't Hooft anomaly of $T_{n_F}$ directly. By gauging $\bZ_m$ symmetry we have:
\begin{equation}
Z_{T_{n_F}/\bZ_m} [\S_g]= \frac{1}{m^g}\sum_{(a_i,b_i)} \epsilon(a_i,b_i) Z_{{n_F}}[\S_g,(a_i,b_i)].
\end{equation}
 For $Z_{{n_F}/\bZ_m}[\S_g]$ to be well defined, each $Z_{{n_F}}[\S_g,(a_i,b_i)]$ needs to be well defined up to an overall constant. This means that, for a background-preserving large diffeomorphism 
\begin{equation}
 M  \begin{bmatrix} a \\[2pt] b \end{bmatrix}= \begin{bmatrix} a \\[2pt] b  \end{bmatrix} \mod \bZ\,, ~~~~~ M \in Sp(2g;\bZ)\,,
 \end{equation}
the quotient
\begin{equation}
\label{eq:partition-quotient}
\frac{M\cdot Z_{{n_F}}[\S_g,(a_i,b_i)]}{Z_{{n_F}}[\S_g,(a_i,b_i)]}
\end{equation}
must be a constant independent of $(a_i,b_i)$. The obstruction to this well-definedness is precisely given by an anomaly. This identity should be interpreted formally, when there is a zero modes in the presence of the $\bZ_m$ background on $\S_g$.  

\subsection{Anomalies from fermion partition functions on $\S_g$}
We now apply the modular transformation law of the theta function to the product formula \eqref{eq:many-fermion-partition-sg}. The resulting phase determines whether the orbifold partition function is well defined in the presence of a flat $\bZ_m$ background via \eqref{eq:partition-quotient}. We first need the $Sp(2g;\bZ)$ transformation properties of \eqref{eq:many-fermion-partition-sg} \cite{igusa2012theta}. For
\begin{equation}
    M =   \begin{pmatrix}
        A&B \\
        C & D
    \end{pmatrix} \in Sp(2g;\bZ)\,,    
\end{equation}
we have
\begin{equation}
    \label{eq:theta-modular-transform}
    \vartheta\!\begin{bmatrix} \tilde{a} \\ \tilde{b} \end{bmatrix} (\tilde{\Omega})
    = \varepsilon(M) e^{-i\pi \Phi(a,b,\Omega)} \det(C\Omega + D)^{1/2}
      \vartheta\!\begin{bmatrix} a \\ b \end{bmatrix} (\Omega)
\end{equation}
where
\begin{equation}
    \label{eq:theta-modular-transform-2}
     \begin{bmatrix} \tilde{a} \\[2pt] \tilde{b} \end{bmatrix}
    = 
    \begin{pmatrix}
        D & -C \\
        -B & A
    \end{pmatrix}
    \begin{bmatrix} a \\[2pt] b \end{bmatrix}
    + \frac{1}{2}
    \begin{bmatrix}
        (CD^{T})_d \\
        (AB^{T})_d
    \end{bmatrix},  
    \end{equation}
   \vspace{0.1cm}
    \begin{equation*}
 \tilde{\Omega} = (A\Omega +B)\cdot (C\Omega+D)^{-1}, 
   \end{equation*}
    \begin{equation*}
    \Phi(a,b,\Omega)
    = [a D^{T} B a + b C^{T} A b]
      - [2 a B^{T} C b + (a D^{T} - b C^{T})(AB^{T})_d]
      \end{equation*}
 Here $(CD^{T})_d = (\sum_{i=1,..,g}C_{1,i}D_{1,i},..,\sum_{i=1,..,g}C_{g,i}D_{g,i})^T$ and  $\varepsilon(M)$ is an eighth root of unity only depends on $M$. The condition \eqref{eq:partition-quotient} now becomes
\begin{equation}
\label{eq:partition-anomaly}
    \begin{split}
    \sum_{l=1}^{n_F} \Phi(q_{l}a,-q_{l}b,\Omega) &= \sum_{l=1}^{n_F} q_{l}^2[a D^{T} B a + b C^{T} A b]\\
      &- \sum_{l=1}^{n_F}[-2q_{l}^2  a B^{T} C b + q_l(a D^{T} + b C^{T})(AB^{T})_d] = 0 \mod 2.
      \end{split}
\end{equation}
\subsubsection{$\bZ_m = \bZ_{2}^F$}
\label{subsec:gso-anomaly-partition}
The analysis here is essentially the same as in \cite{Alvarez-Gaume:1986rcs}. In this case, $q_l =1, l=1,..,n_F$. The $2^{2g}$ $\bZ_{2}^F$ backgrounds $(a_i,b_i)$ on $\S_g$ can be understood as different spin structures for chiral fermions. They come in two sets up to diffeomorphism:
\begin{itemize}
\item The odd spin structure with $\sum_{i=1}^g 4a_ib_i$ odd. Via diffeomorphism transformation they can be set to $(a_1 =1/2,0,0..,b_1 =1/2,0,..,0)$. The Dirac operator of fermions with these spin structures generically has zero modes and $\vartheta\begin{bmatrix}
a \\[.1pt]
-b 
\end{bmatrix}(z=0|\Omega)  =0$ as a result.
\item The even spin structure with  $\sum_{i=1}^g 4a_ib_i$ even. Via diffeomorphism transformation they can be set to $(0,0,0..,0,0,..,0)$. The Dirac operator of fermions with these spin structures generically has no zero modes. 
\end{itemize}
As $Z_{T_{n_F}}[\S_g,(a_i,b_i)] = 0$ for odd spin structures, we focus on the case of even spin structures, transformed to $(0,0,0..,0,0,..,0)$.
Using \eqref{eq:theta-modular-transform-2} we see that $M\in Sp(2g;\bZ)$ preserves this spin structure if 
\begin{equation}
\label{eq:preserve-even-spin}
(CD^{T})_d, ~        (AB^{T})_d \in 2\bZ.
\end{equation}
For $\bZ_2^F$, condition \eqref{eq:partition-anomaly} becomes
\begin{equation}
 n_F [a D^{T} B a + b C^{T} A b]
      - n_F[-2   a B^{T} C b + (a D^{T} + b C^{T})(AB^{T})_d] = 0 \mod 2\,.
\end{equation}
As $a, b \in \frac{\bZ}{2}$, $n_F \in 8\bZ$ automatically satisfies this requirement. When $n_F \in 4\bZ$, we have $n_F[-2   a B^{T} C b + (a D^{T} + b C^{T})(AB^{T})_d] = 0 \mod 2$ automatically. It is non trivial, but can be checked that\footnote{This is tied with ``backfiring" bosonization \cite{BoyleSmith:2024qgx}. We will discuss this later.}
\begin{equation}
n_F [a D^{T} B a + b C^{T} A b] = 0 \mod 2
\end{equation}
 \eqref{eq:preserve-even-spin} holds.   

To conclude, \eqref{eq:partition-anomaly} for $\bZ_2^F$ symmetry of $T_{n_F}$ requires
\begin{equation}
n_F \in 4\bZ\,,
\end{equation}
which is the same as the anomaly-free condition \eqref{eq:anom-free-general-fermions} from bordism perspective discussed in Section \ref{sec:fermanom}. Moreover, the partition function of $T_{n_F=4m}$ after gauging $\bZ_2^F$ on $\S_g$ is
\begin{equation}
Z_{{n_F}/\bZ_2^F} [\S_g]= \frac{1}{2^g}\sum_{(a_i,b_i: 4\sum_{i}a_ib_i\in 2\bZ)}   \cdot \prod_{l=1}^{4m}\epsilon(a_i,b_i)\vartheta\begin{bmatrix}
a \\[.1pt]
 b 
\end{bmatrix} (z=0|\Omega)  
\end{equation}
Here $\epsilon(a_i,b_i)$ are some well defined factors to make the partition function modular invariant.

\subsubsection{$\bZ_m \times \bZ_2^F$}
After discussing the $\bZ_2^F$ symmetry, we can move on to the $\bZ_m \times \bZ_2^F$ symmetries. Requiring $\bZ_2^F$ symmetry is anomaly free gives $n_F \in 4\bZ$. The analysis of condition \eqref{eq:partition-anomaly} for $m$ even is done in \cite{Freed:1987qk}, it is shown that they are equivalent to the anomaly free conditions derived in last section using bordism groups. For $m$ odd, the same analysis can be done similarly. 

Another way to see the conditions for $m$ odd is the following. As we have $n_F \in 4\bZ$ copies of fermions, we can assign a ('t Hooft) anomaly free ``$\bZ_2$" symmetry to $T_{n_F}$ by giving each fermion $\bZ_2$ charge $1$. Since $m$ is odd, $\bZ_{m}\times \bZ_2 = \bZ_{2m}$. With the anomaly free "$\bZ_2$" symmetry, we can study the anomaly of $\bZ_{2m}$ symmetry, where each fermion has $\bZ_{2m}$ charge $2q_{l},l=1,...,n_F$.

The anomaly-free condition for $\bZ_{2m}$ implies anomaly-free condition for $\bZ_m$ symmetry. Since $2m$ is even, the analysis from partition function \cite{Freed:1987qk} gives
\begin{equation}
    \sum_{i=1}^{n_F} 4 q_l^2 = 0 \mod 4m, \quad \sum_{i=1}^{n_F} 2 q_l = 0 \mod2  \quad\Rightarrow\quad \sum_{i=1}^{n_F} q_l^2 = 0 \mod m\,,
\end{equation}
which is equivalent to \eqref{eq:full-anomaly}. From direct analysis of the partition functions of $T_{n_F}$ on $\S_{g\geq 1}$, we see the anomaly free conditions from bordism group perspective guarantees the partitions function of gauged theory is well defined.

\subsubsection{One loop partition of fermions: $\S_g$ with $g=1$}
We now discuss the well-known genus-one example of the partition function of $T_{n_F}$. 

For $g =1$, $\Omega$ is a complex number $\tau$ with $\text{Im}(\tau)>0$. We have four spin structures on $\S_{g=1}$ ($\bZ_2^F$ backgrounds) and the corresponding partition functions are
\begin{equation}
    \label{eq:genus-one-weyl-partition}
    \begin{split}
        & Z_{{n_F}}[NS,NS]   = \frac{\vartheta\begin{bmatrix}0 \\[.1pt]0\end{bmatrix}^{n_F}(z=0|\tau) }{\eta^{n_F}(q)}\,,\\
        &Z_{{n_F}}[NS,R]   = \frac{\vartheta\begin{bmatrix}0 \\[.1pt]\tfrac12\end{bmatrix}^{n_F}(z=0|\tau) }{\eta^{n_F}(q)}\,,\\
        &Z_{{n_F}}[R,NS]   = \frac{\vartheta\begin{bmatrix}\tfrac12\\[.1pt]0\end{bmatrix}^{n_F}(z=0|\tau) }{\eta^{n_F}(q)}\,,\\
        &Z_{{n_F}}[R,R]    = \frac{\vartheta\begin{bmatrix}\tfrac12\\[.1pt]\tfrac12\end{bmatrix}^{n_F}(z=0|\tau) }{\eta^{n_F}(q)}=0\,.
    \end{split}
\end{equation}
where $q = \exp(2\pi i \tau)$. These transform under the two $SL(2;\bZ)$ generators $S:~\tau \to -1/\tau$ and $T:~\tau\to \tau+1$ as \footnote{up to an overall consant $e^{-\frac{n_F 2\pi i}{24}}$ related to perturbative gravitational anomaly.}
\begin{align}
    &S: ~~~~~Z_{{n_F}}[R,NS] \leftrightarrow  Z_{{n_F}}[NS,R], ~~~~~Z_{{n_F}}[NS,NS] \to Z_{{n_F}}[NS,NS]\,,\\
    &T: ~~~~~ Z_{{n_F}}[NS,NS]\leftrightarrow   Z_{{n_F}}[NS,R], ~~~ Z_{{n_F}}[R,NS] \to i^{n_F/2}Z_{{n_F}}[R,NS]\,.
\end{align}
The prefactor $i^{n_F/2}$ in the second line is precisely what makes the partition function after gauging $\bZ_2^F$  be modular invariant only when $n_F \in 4\bZ$. In this case, the partition function takes the form
\begin{equation}
\label{eq:one-loop-gauging-z2f}
    Z_{{n_F}/\bZ_2^F} = \frac{1}{2} ((-1)^{n_F/4}Z_{{n_F}}[NS,NS]+  Z_{{n_F}}[R,NS]+ Z_{{n_F}}[NS,R])\,, 
\end{equation}
and the prefactor $(-1)^{n_F/4}$ distinguishes the two cases $n_F \in 8\bZ$ and $n_F =4 \mod 8\bZ$. This distinction is closely tied to the phenomenon of ``backfiring bosonization" \cite{BoyleSmith:2024qgx}: $T_{n_F}/\bZ_2^F$ depends on spin structure when $n_F =4 \mod 8\bZ$ and does not depend on it when $n_F = 0 \mod 8$. It would be interesting to investigate the generalization on higher genus surfaces.

\section{Bosonization}
\label{sec:bosonization}
The heterotic worldsheet theory $\cT_\text{HE}=\cT/G_\text{GSO}$ also admits a bosonic formulation in which the left-moving fermions $\lambda_L^A$ and $\lambda_L^{A'}$ are replaced by scalar fields $X_L^A$ and $X_L^{A'}$, yielding the $E_8\oplus E_8$ lattice conformal field theory. This raises a natural question: how do the anomaly constraints discussed in the fermionic description appear in the bosonic one? One possibility would be to develop a direct anomaly analysis for chiral scalar fields, but such a formalism is not yet well developed to our knowledge.\footnote{When the automorphism group is abelian, it is expected that the level-matching condition coincides with the anomaly-free condition \cite{Vafa:1986wx,Harvey:1987da,Baykara:2024vss}. Matching the level-matching condition with the anomaly-free condition in the fermionic theory $\cT$ provides a proof of this statement in the present setting.} Instead, we analyze these symmetries by transporting them to the fermionic theory $\cT$, where the anomaly analysis of Section \ref{sec:fermanom} applies directly. The goal of this section is to make the matching between the bosonic and fermionic descriptions explicit for a large class of $\mathfrak{e}_8\oplus\mathfrak{e}_8$ inner automorphisms.

The main claim of this section is the following. Let \(g_w\) be a finite-order inner
automorphism of the \(E_8\) lattice CFT represented by a shift vector
\(w\in \Gamma_{E_8}\otimes \mathbb Q\), and let \(m\) be its order. Define
\(Q=mw\in \Gamma_{E_8}\). Suppose that the symmetry survives
fermionization, so that it defines a symmetry of \(T_F\). Then the bosonic anomaly
class of \(g_w\) is equal to the fermionic anomaly class computed from the charges
of the eight Weyl fermions. Explicitly,
\[
[Q^2]_{N}
=
\left[\sum_{A=1}^8 q_A^2\right]_{N},
\qquad
N=
\begin{cases}
m, & m\ \mathrm{odd},\\
2m, & m\ \mathrm{even}.
\end{cases}
\]
Thus the bosonic level-matching condition is equivalent to the fermionic bordism
anomaly cancellation condition for this class of symmetries.

\subsection{Bosonization and fermionization}
\label{subsec:bosonization-fermionization}

To compare the anomaly discussion of Section~2 with the bosonic \(E_8\) lattice description, it is useful to isolate a single left-moving block. Let \(T_F\) denote the theory of eight left-moving Weyl fermions \(\lambda_L^A\), \(A=1,\dots,8\), and let
\begin{equation}
\label{eq:bosonization}
    T_B \equiv T_F/\bZ_2^F
\end{equation}
be the theory obtained by gauging fermion parity. The point of view throughout this section is that \(T_F\) and \(T_B\) are different theories in the modern sense: \(T_F\) is a fermionic theory, hence it depends on a choice of spin structure, whereas \(T_B\) is bosonic and does not. At the same time, the bosonic theory \(T_B\) admits both a free-fermion realization and a lattice-boson realization. Accordingly, the words \emph{bosonization} and \emph{fermionization} are sometimes used in two different ways in the literature: either for the equivalence between two presentations of the same bosonic CFT, or for the operations relating a fermionic theory to a bosonic one. We will use the latter meaning.

With this convention, bosonization is the passage from \(T_F\) to \(T_B\) in \eqref{eq:bosonization}, while fermionization is
\begin{equation}
\label{eq:fermionization}
    T_F = (T_B \times \mathrm{Arf})/\bZ_{2,b}\,.
\end{equation}
Here \(\bZ_{2,b}\) is the quantum symmetry of the \(\bZ_2^F\) orbifold, and the extra Arf theory is included so that the result is again fermionic rather than bosonic. In the bosonic presentation, \(T_B\) is the \(E_8\) lattice conformal field theory of eight chiral bosons.

\subsubsection{The lattice dictionary}
\label{subsec:lattice-dictionary}

We now spell out the relation between the bosonic and fermionic descriptions. The key point is that there are two distinct lattices in the story.

First, the bosonic theory \(T_B\) is the chiral lattice CFT associated with the \(E_8\) root lattice \(\Gamma_{E_8}\). A convenient basis for \(\Gamma_{E_8}\) is given by the simple roots \(\alpha_i\), \(i=1,\dots,8\), whose components in an orthonormal frame are
\begin{equation}
\label{eq:e8-root-lattice}
    \begin{pmatrix}
        \alpha_{1} \\
        \alpha_{2} \\
        \alpha_{3} \\
        \alpha_{4} \\
        \alpha_{5} \\
        \alpha_{6} \\
        \alpha_{7} \\
        \alpha_{8}
    \end{pmatrix}
    =
    \begin{pmatrix}
        1 & -1 & 0 & 0 & 0 & 0 & 0 & 0 \\
        0 & 1 & -1 & 0 & 0 & 0 & 0 & 0 \\
        0 & 0 & 1 & -1 & 0 & 0 & 0 & 0 \\
        0 & 0 & 0 & 1 & -1 & 0 & 0 & 0 \\
        0 & 0 & 0 & 0 & 1 & -1 & 0 & 0 \\
        0 & 0 & 0 & 0 & 0 & 1 & -1 & 0 \\
        0 & 0 & 0 & 0 & 0 & 0 & 1 & 1 \\
        -\frac{1}{2} & -\frac{1}{2} & -\frac{1}{2} & -\frac{1}{2} &
        -\frac{1}{2} & \frac{1}{2} & \frac{1}{2} & -\frac{1}{2}
    \end{pmatrix}.
\end{equation}
Using this basis, the bosonic theory is described by eight chiral scalars \(\phi_i\sim \phi_i+2\pi\), and the vertex operators are labeled by lattice vectors
\begin{equation}
\label{eq:tb-vertex-operators}
    \ell=\sum_{i=1}^8 \ell_i \alpha_i \in \Gamma_{E_8},
    \qquad
    V_\ell = :e^{\,i\sum_{i=1}^8 \ell_i \phi_i}: \, .
\end{equation}
Equivalently, if one writes \(\ell\) in the simple-root basis, the coefficients \(\ell_i\) are integers.

Second, before the GSO projection the free-fermion theory \(T_F\) is naturally described by the orthonormal lattice
\begin{equation}
\label{eq:z8-lattice}
    \bZ^8 = \bigoplus_{A=1}^8 \bZ e_A,
    \qquad
    e_A=(\delta_{1A},\dots,\delta_{8A}),
\end{equation}
where \(e_A\) is associated with the Weyl fermion \(\lambda_L^A\). This is not the same lattice as \(\Gamma_{E_8}\), although both sit naturally inside the same ambient Euclidean space \(\bR^8\). The matrix in \eqref{eq:e8-root-lattice} should therefore be regarded only as a change of basis inside \(\bR^8\): it expresses the \(E_8\) simple roots in the orthonormal frame adapted to the fermions. It does \emph{not} identify \(\Gamma_{E_8}\) with \(\bZ^8\).

The passage from \(T_F\) to \(T_B\) is best expressed in lattice language. The charge lattice of the free-fermion theory decomposes as
\begin{equation}
\label{eq:z8-decomposition}
    \bZ^8 = D_8 \cup (D_8+v),
\end{equation}
where \(D_8\) is the \(SO(16)\) root lattice and \(v\) is a vector weight. The untwisted \(\bZ_2^F\)-even operators have charges in \(D_8\), while the untwisted \(\bZ_2^F\)-odd operators have charges in \(D_8+v\). Gauging \(\bZ_2^F\) therefore projects the untwisted sector down to the \(D_8\) sublattice. The twisted sector contributes the spin fields, whose charges lie in
\begin{equation}
\label{eq:spinor-coset}
    D_8+s,
\end{equation}
with \(s\) a chiral spinor weight. Altogether the orbifold theory has charge lattice
\begin{equation}
\label{eq:e8-from-d8}
    D_8 \cup (D_8+s)=\Gamma_{E_8}.
\end{equation}
This is the lattice-level statement that
\begin{equation}
\label{eq:tb-equals-tf-gso}
    T_B = T_F/\bZ_2^F
\end{equation}
is the \(E_8\) lattice CFT.

It is useful to keep in mind a concrete example. Consider in \(T_B\) the vertex operator associated with the lattice vector \(\ell=w_1\alpha_1\), with \(w_1\in \bZ_{>0}\). In the orthonormal basis \(e_A\), the same vector is
\begin{equation}
\label{eq:alpha1-orthonormal}
    w_1\alpha_1=(w_1,-w_1,0,\dots,0).
\end{equation}
Thus the corresponding bosonic vertex operator
\begin{equation}
\label{eq:bosonic-vertex-example}
    :e^{i w_1 \phi_1}:
\end{equation}
is represented in the free-fermion description schematically, by the fermionic bilinear operator:
\begin{equation}
\label{eq:fermionic-vertex-example}
    :(\lambda_L^1)^{w_1}(\bar\lambda_L^2)^{w_1}: \, .
\end{equation}
More generally, the same state may be described either as a vertex operator labeled by a vector in \(\Gamma_{E_8}\) or, in the fermionic realization, by an operator whose \(U(1)^8\) charges are read in the orthonormal basis \(e_A\).

A nontrivial consequence of the equivalence between \(T_B\) and \(T_F/\bZ_2^F\) is that any symmetry preserved by the bosonization/fermionization procedure must have the same anomaly in both descriptions. In the rest of this section we make this statement explicit for a large class of finite-order inner automorphisms of the \(E_8\) lattice theory. The strategy is simple: start with a symmetry of \(T_B\), determine when it lifts to a genuine symmetry of \(T_F\), and then compare the resulting anomaly classes on the two sides.

\subsection{Inner automorphisms and anomaly matching}
As we have said, we are interested in automorphisms of the $\mathfrak{e}_8 \oplus \mathfrak{e}_8$ current algebra. These can be either \textit{inner} or \textit{outer} automorphisms, the latter of which are characterized by involving the trading of the two $\mathfrak{e}_8$ subalgebras. Outer automorphisms do not commute with the GSO projection when transported to $T_F \times T_F$, hence they fall outside of the class of symmetries considered in Section 2; we come back to this problem at the end of this section. On the other hand, to study inner automorphisms it suffices to consider one copy of $T_B$, since the results extend trivially to $T_B \times T_B$.

\subsubsection{Finite-order inner automorphisms and Kac's algorithm}

The symmetries of interest are finite-order inner automorphisms of the $E_8$ current algebra. Such automorphisms are conjugate to elements of a maximal torus of $E_8$, so their action on a momentum eigenstate $|P\rangle$ with $P \in \Gamma_{E_8}$ can be written as
\begin{equation}
\label{eq:g-action-shift}
    g(w)|P\rangle = e^{2\pi i\, w \cdot P}\, |P\rangle,
\end{equation}
where $w \in \Gamma_{E_8}\otimes \mathbb{Q}$ is the corresponding shift vector. Two shift vectors related by the Weyl group define conjugate automorphisms, and Kac's theorem \cite{kac1990infinite} gives a convenient choice of representative inside a fundamental Weyl chamber. Namely, if $g$ has order $m$, then there exists a set of non-negative integers $s_0,\dots,s_8$, relatively prime as a set, satisfying
\begin{equation}
\label{eq:kac-constraint}
    \sum_{a=0}^8 s_a \kappa_a = m,
\end{equation}
where $\kappa_a$ are the marks of the affine $E_8$ Dynkin diagram, such that the shift vector can be written as
\begin{equation}
\label{eq:shift-vector}
    w = \frac{1}{m}\sum_{a=1}^8 s_a \mu_a,
\end{equation}
with $\mu_a$ the fundamental weights of $E_8$. The marks read $\kappa_a = 2,3,4,5,6,4,2,3$ in the ordering of \eqref{eq:e8-root-lattice}, and $\kappa_0 \equiv 1$ for the lowest root.

In the bosonic realization of $T_B$, the generator $g$ of the $\mathbb{Z}_m$ symmetry acts on the chiral bosons as
\begin{equation}
\label{eq:g-action-bosons}
    g \cdot \phi_i = \phi_i + 2\pi (w,\alpha_i)_{\Gamma_{E_8}} .
\end{equation}

\subsubsection{Anomaly in $T_B$}

At a minimum, gauging $\bZ_m$ symmetry (generated by $g$ \eqref{eq:g-action-bosons}) consistently in $T_B$ requires the one-loop partition function of the orbifold theory to be modular invariant. Inserting $g(w)$ in the trace over the untwisted sector and applying the $S$ and $T^m$ modular transformations, one finds
\begin{equation}
\label{eq:tb-partition-transform}
    Z_{1,g}[T_B]
    = \frac{1}{\eta^8}\sum_{P \in \Gamma_{E_8}} e^{2\pi i\, w \cdot P} q^{\tfrac12 P^2}
    \qquad \Longrightarrow \qquad
    Z_{g,g^m}[T_B]
    = \frac{1}{\eta^8}\sum_{P \in \Gamma_{E_8}} q^{\tfrac12 (P+w)^2} e^{\pi i m (P+w)^2}.
\end{equation}
One-loop modular invariance therefore requires
\begin{equation}
    m(P+w)^2 \in 2\mathbb{Z}
\end{equation}
for all $P \in \Gamma_{E_8}$. Since $m w \in \Gamma_{E_8}$, this is equivalent to
\begin{equation}
\label{eq:tb-anomaly}
    m \frac{w^2}{2}
    = \frac{1}{2m}\sum_{a,b=1}^8 s_a s_b\, h^{-1,ab}_{E_8}
    \in \mathbb{Z},
\end{equation}
where $h^{-1,ab}_{E_8}$ is the inverse of the Cartan matrix of $E_8$. This is the usual level-matching condition
\begin{equation}
\label{eq:level-matching}
    m E_{0,g} \in \mathbb{Z},
\end{equation}
where $E_{0,g} = \frac12 w^2$ is the zero-point energy in the $g$-twisted sector. See also \cite{Chang:2018iay,BoyleSmith:2023xkd} for the interpretation of the same condition from topological defect lines.

When
\begin{equation}
    m \frac{w^2}{2} \not\equiv 0 \pmod{\mathbb{Z}},
\end{equation}
we interpret this nonzero class as the 't Hooft anomaly of the corresponding $\mathbb{Z}_m$ symmetry \cite{Chang:2018iay}. Indeed, if we regard $Z_{g;1}[T_B]$ as the one-loop partition function of $T_B$ in the background $(g;1)$, then $T^m$ preserves this background,
\begin{equation}
    T^m : (g;1) \longrightarrow (g;g^m) = (g;1),
\end{equation}
and the resulting phase ambiguity is
\begin{equation}
\label{eq:level-matching-anomaly}
    \frac{T^m \cdot Z_{g;1}[T_B]}{Z_{g;1}[T_B]}
    = e^{2\pi i\, m \frac{w^2}{2}}.
\end{equation}
Thus the anomaly class is represented by $m \frac{w^2}{2} \bmod \mathbb{Z}$.

It is convenient to define
\begin{equation}
\label{eq:Q-def}
    Q \equiv m w \in \Gamma_{E_8}.
\end{equation}
Then \eqref{eq:tb-anomaly} is equivalent to
\begin{equation}
    Q^2 \equiv 0 \pmod{2m}.
\end{equation}
When $m$ is odd, this condition is equivalent to $Q^2 \equiv 0 \pmod m$, since
$\Gamma_{E_8}$ is even and hence $Q^2 \in 2\mathbb{Z}$ automatically. The bosonic
anomaly-free condition is therefore
\begin{equation}
\label{eq:bosonic-anomaly-free}
    Q^2 =
    \begin{cases}
        0 \pmod{2m}, & m \text{ even},\\[4pt]
        0 \pmod m, & m \text{ odd}.
    \end{cases}
\end{equation}
These conditions are very similar to the anomaly cancellation conditions in $T_F$, cf. \eqref{eq:anom-free-general-fermions}. Moreover, the bosonic anomaly class $m \frac{w^2}{2} \bmod \mathbb{Z}$ should be reproduced on the $T_F$ side whenever the corresponding $\mathbb{Z}_m$ symmetry survives the bosonization/fermionization procedure.

The following example is instructive.
The theory $T_F$ can be obtained by gauging the quantum symmetry $\mathbb{Z}_{2,b}$ of
$T_B$, see \eqref{eq:fermionization}. For this gauging to be consistent,
$\mathbb{Z}_{2,b}$ itself must be free of anomaly. In this case the corresponding shift
vector is
\begin{equation}
\label{eq:z2b-shift}
    w = \frac{\mu_7}{2},
    \qquad
    Q = 2w = \mu_7 .
\end{equation}
Then
\begin{equation}
    Q^2 \equiv 0 \pmod 4,
\end{equation}
so the anomaly-free condition \eqref{eq:bosonic-anomaly-free} is satisfied. See \cite{BoyleSmith:2023xkd} for more details on this example.

\subsubsection{Lifting bosonic symmetries to $T_F$}
\label{subsec:anomalies-in-tf}

To identify the corresponding $\mathbb{Z}_m$ action in $T_F$, recall that the eight Weyl fermions $\lambda_L^A$ furnish sixteen Majorana--Weyl fermions transforming in the fundamental representation of $SO(16)$. More generally, a state of weight $\mu$ transforms under $g$ as
\begin{equation}
    g: |\mu\rangle \mapsto e^{2\pi i\, w\cdot \mu} |\mu\rangle .
\end{equation}
For the fundamental representation we may take $\mu_A=e_A$, so the $\mathbb{Z}_m$ charge of $\lambda_L^A$ is
\begin{equation}
    q_A = m\, w\cdot e_A .
\end{equation}
Using the notation introduced in \eqref{eq:Q-def}, the $\mathbb{Z}_m$ charges of the
fermions are obtained by reading the components of $Q$ in the orthonormal basis $e_A$
associated with the fermions. Equivalently,
\begin{equation}
\label{eq:qA-from-Q}
    q_A = Q\cdot e_A .
\end{equation}
Thus $Q$ is naturally a vector in the bosonic lattice, viewed inside the ambient space
$\mathbb{R}^8$, while the fermionic charges are its components in the orthonormal basis.

Now, recall that $T_F$ is obtained from $T_B$ by gauging $\mathbb{Z}_{2,b}$. For the symmetry generated by $w$ to survive fermionization, the charges $q_A$ must be well defined in the fermionic theory. Equivalently, $Q$ must have integral components in the orthonormal basis. Since
\begin{equation}
    \mathbb{Z}^8 = D_8 \cup (D_8+v),
\end{equation}
this means that for even $m$ we must require
\begin{equation}
\label{eq:Q-in-D8}
    (Q,v) \in \bZ.
\end{equation}
Indeed, for $x\in D_8$ we automatically have $(Q,x)\in\mathbb{Z}$ because $Q\in \Gamma_{E_8}$, but for the charges of the elementary fermions to be well defined we also need $(Q,v)\in\mathbb{Z}$. Since $v\notin \Gamma_{E_8}$, this is a nontrivial condition. Writing $v$ in the simple-root basis \eqref{eq:e8-root-lattice},
\begin{equation}
    v = \sum_{i=1}^6 i\,\alpha_i + \frac{7}{2}(\alpha_8-\alpha_6),
\end{equation}
and using $Q=mw$ with $w$ given by \eqref{eq:shift-vector}, we obtain
\begin{equation}
\label{eq:qv-even-m}
    (Q,v)=\frac{7}{2}(s_8-s_6)\in \mathbb{Z}
    \qquad \Longrightarrow \qquad
    s_8-s_6\in 2\mathbb{Z}.
\end{equation}
This is the condition that the bosonic symmetry lift to a symmetry of $T_F$ when $m$ is even.

When $m$ is odd, the situation is simpler. Since $2$ is invertible modulo $m$, if $g$ generates $\mathbb{Z}_m$ then so does $g^2$. Replacing $g$ by $g^2$ replaces $Q$ by $2Q$, and $2Q$ always lies in $D_8$. Thus for odd $m$ there is no additional integrality condition analogous to \eqref{eq:qv-even-m}.

There is one further subtlety when $m=4k$. In that case we must also require that $\mathbb{Z}_{2,b}$ is not the order-two subgroup of the bosonic $\mathbb{Z}_m$. Otherwise, after gauging $\mathbb{Z}_{2,b}$ the resulting symmetry in $T_F$ is not the original $\mathbb{Z}_m$, but rather a quotient or extension involving fermion parity. In particular, if
\begin{equation}
    2k\, w = \frac{\mu_7}{2},
\end{equation}
then the order-two subgroup generated by $g^{2k}$ is precisely $\mathbb{Z}_{2,b}$, and one has an exact sequence
\begin{equation}\label{extension m=4k}
    0 \to \mathbb{Z}_{2,b} \to \mathbb{Z}_m \to \mathbb{Z}_{2k} \to 0 .
\end{equation}
After gauging $\mathbb{Z}_{2,b}$, the induced symmetry on $T_F$ is  $\bZ_{2k}\times \bZ_{2}^F$ with a mixed 't Hooft anomaly given by non trivial extension \eqref{extension m=4k} \cite{Tachikawa:2017gyf}. One class of examples satisfies this condition arises for shift vectors \eqref{eq:shift-vector} with $s_7=0$.

We will therefore restrict attention to bosonic $\mathbb{Z}_m$ symmetries satisfying the following conditions:
\begin{equation}
\label{eq:lift-conditions}
\begin{cases}
m \text{ odd},\\[4pt]
m \text{ even}: \text{\eqref{eq:Q-in-D8} holds, and } \mathbb{Z}_{2,b} \not\subset \mathbb{Z}_m \text{ when } 4\mid m.
\end{cases}
\end{equation}
For such symmetries we can compare the anomaly in $T_F$ with the bosonic anomaly discussed above.

\subsubsection{Fermionic anomaly and anomaly matching}

Assume first that $m$ is even. Since $T_F$ has a perturbative gravitational anomaly, we temporarily add right-moving Weyl fermions $\lambda_R^A$ neutral under $\mathbb{Z}_m$ in order to isolate the discrete anomaly. Consider the one-loop partition function with NS--NS spin structure and background $(g;1)$, denoted $Z_{(NS,NS),(g;1)}[T_F]$. Then $T^m$ preserves this background and we find
\begin{equation}
\label{eq:level-matching-in-tf}
\frac{T^m\cdot Z_{(NS,NS),(g;1)}[T_F]}{Z_{(NS,NS),(g;1)}[T_F]}
=
e^{2\pi i \sum_{A=1}^8 \frac{q_A^2}{2m}} .
\end{equation}
Hence the fermionic anomaly class is
\begin{equation}
\label{eq:level-matching-fermions}
    \sum_{A=1}^8 q_A^2 \pmod{2m}.
\end{equation}
On the other hand, the bosonic anomaly class is $Q^2 \pmod{2m}$. Since the basis $e_A$ is orthonormal,
\begin{equation}
\label{eq:anomaly-matching-even-m}
    Q^2 = \sum_{A=1}^8 q_A^2 .
\end{equation}
Thus the bosonic and fermionic anomaly classes agree. The symmetry is anomaly-free precisely when this common class vanishes modulo $2m$. In particular, the fermionic anomaly-free condition obtained from \eqref{eq:level-matching-in-tf} is exactly the condition derived in Section~\ref{sec:fermanom}.

Combining fermionization \eqref{eq:fermionization}, the anomaly matching \eqref{eq:anomaly-matching-even-m}, and the bordism analysis of Section~\ref{sec:fermanom}, we conclude that for this class of symmetries the one-loop level-matching condition in $T_B$ is sufficient to guarantee consistency after gauging $\mathbb{Z}_m$. It would be interesting to extend this logic to more general chiral-boson systems.

Now suppose that $m$ is odd. As explained above, we may replace the generator $g$ by $g^2$, which is again a generator of $\mathbb{Z}_m$, and work with $2Q\in D_8$. The same analysis then gives the fermionic anomaly class\footnote{For simplicity we write the formulas in terms of the original $Q$ and $q_A$.}
\begin{equation}
    \sum_{A=1}^8 q_A^2 \pmod m,
\end{equation}
while the bosonic anomaly class is $Q^2 \pmod m$. Since again
\begin{equation}
\label{eq:anomaly-matching-odd-m}
    Q^2 = \sum_{A=1}^8 q_A^2,
\end{equation}
the two anomaly classes agree modulo $m$. The symmetry is anomaly-free precisely when this class vanishes, in agreement with the condition derived in Section~\ref{sec:fermanom}.

\paragraph{Remark}
Another way to see that a $\mathbb{Z}_m$ symmetry with $m$ odd is compatible with
gauging $\mathbb{Z}_{2,b}$ is to note that the symmetry group in $T_B$ is
$\mathbb{Z}_{2,b}\times \mathbb{Z}_m$. If there were a mixed anomaly between these
two factors, the corresponding symmetry in $T_F$ would have to be a nontrivial
extension of $\mathbb{Z}_m$ by $\mathbb{Z}_2^F$ \cite{Tachikawa:2017gyf}, which is
impossible for odd $m$.

A related point of view is to combine the generator of $\mathbb{Z}_m$ with
$\mathbb{Z}_{2,b}$ and regard the symmetry in $T_B$ as a $\mathbb{Z}_{2m}$ symmetry.
After fermionization, this becomes the corresponding $\mathbb{Z}_{2m}$ symmetry in
$T_F$, with $\mathbb{Z}_2^F$ as its order-two subgroup. Since $\mathbb{Z}_{2,b}$ and
$\mathbb{Z}_2^F$ are anomaly-free, the potentially anomalous part is the same in both
descriptions, and the odd-$m$ case may be viewed as a reformulation of the even-order
analysis applied to $\mathbb{Z}_{2m}$.

We conclude with two examples illustrating the general matching statement. The first
has a nontrivial anomaly, while the second is anomaly-free. The corresponding $\bZ_m$
 charges in $T_F$ are read out using the specific $D_8$ sublattice inside $\G_{E_8}$ used in \cite{BoyleSmith:2023xkd}.
\paragraph{A $\mathbb{Z}_4$ example with nontrivial anomaly.}
Consider the shift
\begin{equation}
    w=\frac{\mu_3}{4},
    \qquad
    Q=4w=\mu_3 .
\end{equation}
In the bosonic theory this defines a $\mathbb{Z}_4$ symmetry acting as
\begin{equation}
    \sigma:\ \phi_3 \mapsto \phi_3+\frac14,
    \qquad
    \phi_{i\neq 3}\mapsto \phi_{i\neq 3}.
\end{equation}
This symmetry survives fermionization. In the orthonormal basis, the corresponding
fermion charges are
\begin{equation}
    (q_1,\dots,q_8)=(1,1,1,0,0,0,0,-1).
\end{equation}
The bosonic anomaly class is
\begin{equation}
    Q^2 = h^{-1}_{E_8,33} = 4 \pmod 8,
\end{equation}
while the fermionic anomaly class is
\begin{equation}
    \sum_{A=1}^8 q_A^2 = 4 \pmod 8.
\end{equation}
Thus the same nontrivial anomaly appears in both descriptions.

\paragraph{A $\mathbb{Z}_5$ anomaly-free example.}
Consider instead the shift
\begin{equation}
    w=\frac{\mu_4}{5},
    \qquad
    Q=5w=\mu_4 .
\end{equation}
The corresponding fermion charges are
\begin{equation}
    (q_1,\dots,q_8)=(1,1,1,1,0,0,0,-1).
\end{equation}
In this case the bosonic anomaly class vanishes,
\begin{equation}
    Q^2 = h^{-1}_{E_8,44} = 0 \pmod 5,
\end{equation}
and the fermionic anomaly class vanishes as well,
\begin{equation}
    \sum_{A=1}^8 q_A^2 = 0 \pmod 5.
\end{equation}
So this gives an explicit anomaly-free symmetry for which the two descriptions agree.

\subsection{Outer automorphisms: a brief example}
\label{subsec:outer-automorphisms}
So far we have focused on abelian groups. We now discuss a simple nonabelian example. Consider two copies of the $E_8$ lattice CFT, namely $T_{B,1}\times T_{B,2}$. Starting from the $\bZ_m$ symmetry of $T_B$, we can construct a family of nonabelian symmetry groups by adjoining the $\bZ_{2}$ outer automorphism that exchanges $T_{B,1}$ and $T_{B,2}$:
\begin{equation}\label{eq:outer-automorphism-group}
	(\mathbb{Z}_{m,1} \times \mathbb{Z}_{m,2})\rtimes \mathbb{Z}_{2,\text{out}}\,.
\end{equation}

Via fermionization \eqref{eq:fermionization}, we can use the bordism method to compute the nonperturbative anomaly of this symmetry, provided that the symmetry is not broken by fermionization. Here we limit ourselves to the special case that the symmetry group is $(\bZ_{2}^{F}\times \bZ_{2}^F)\rtimes \bZ_{2}$ the fermionic theory $T_{F,1}\times T_{F,2}$. The anomaly of $(\bZ_{2}^{F}\times \bZ_{2}^F)\rtimes \bZ_{2}$ is valued in \cite{Davighi_2022}: 
\begin{equation}
	\Hom( \Omega_{3d}^{Spin,tor}(B(\mathbb{Z}_{2,out}\ltimes (\mathbb{Z}_{2,1}\times \mathbb{Z}_{2,2})),U(1)) = \mathbb{Z}_8 \oplus \mathbb{Z}_8 \oplus \mathbb{Z}_8.
\end{equation}
Assume the anomaly of MW fermions form a two dimensional representation of $(\bZ_{2}^{F}\times \bZ_{2}^F)\rtimes \bZ_{2}$ is $\Psi \in \bZ_{8}\oplus \bZ_{8}\oplus \bZ_{8}$. Then for $8$ copies of such MW fermions,  $(\bZ_{2}^{F}\times \bZ_{2}^F)\rtimes \bZ_{2}$ is automatically free of anomaly as $8 \Psi=0\in \bZ_{8}\oplus \bZ_{8}\oplus \bZ_{8}$. As a result, for $T_{F,1}\times T_{F,2}$ the symmetry $(\bZ_{2}^{F}\times \bZ_{2}^F)\rtimes \bZ_{2}$ is free of anomaly as it consists of 16 copies of such MW fermions.

As in the $\bZ_m$ examples discussed above, fermionization together with bordism methods can be applied to finite nonabelian groups such as \eqref{eq:outer-automorphism-group}. This should provide all-loop consistency conditions for asymmetric orbifolds with nonabelian orbifold group, and we hope to return to this problem elsewhere.

\section{Conclusion}
\label{sec:conclusion}
In this paper we showed that, for abelian orbifolds of fermionic CFTs relevant to perturbative heterotic string vacua, the usual consistency conditions coincide with the $\bZ_m\times \bZ_2^F$ anomaly-free conditions obtained from the bordism/Dai--Freed viewpoint. In Section \ref{sec:fermanom} we derived these constraints from the anomaly analysis of worldsheet fermions. In Section \ref{sec:theta} we recovered the same conditions directly from partition functions on general Riemann surfaces. In Section \ref{sec:bosonization} we matched the corresponding anomaly data between the fermionic and bosonic descriptions for a large class of symmetries. Taken together, these results explain why the familiar one-loop level-matching conditions are sufficient to ensure consistency to all loop orders in the examples studied here.

There are several natural directions for future work:
\begin{itemize}
    \item Generalize the consistency conditions \eqref{eq:full-anomaly} to non-abelian groups.\footnote{For a collection of bordism groups with non abelian groups and relation to anomalies,  see \cite{Davighi_2022}.} It is shown that conditions in \eqref{eq:full-anomaly} are not enough to guarantee the corresponding orbifold CFT to all loop orders \cite{Freed:1987qk}. This is also expected from the anomaly theory perspective, as the bordism group will have new ingredients and hence gives more constraints. 
    \item Generalize to more general CFTs like the Narain CFT corresponding to lattice $\Gamma_{d,d+16}$. With the full set of anomaly free conditions for general orbifold group $G$ on the fermion side as well as possible bosonization procedure spelled out, it may provide a way to give a complete set of consistency conditions of orbifold Narain CFT to be consistent to all loop orders \cite{Aldazabal:2025zht,Harvey:2017rko}.
    \item Bosonization/Fermionization in higher dimensions, as well as for (2d) chiral bosons. Some  recent investigations on this aspect can be found \cite{BoyleSmith:2024qgx, BoyleSmith:2025duo}. In the modern way of understanding Bosonizaton, which is taken in \cite{BoyleSmith:2024qgx, BoyleSmith:2025duo}, it is understood that the bosonic theory is independent of the spin structure of spacetime by definition. On the other hand, in many cases theory of $2d$ chiral bosons requires a spin structure as input \cite{Witten:1999vg, Eguchi:1986ui,Belov:2005ze}. One way to see it is via AdS/CFT (or generalized CSW/BCFT) duality \cite{Gukov:2004id, Hsieh:2019iba}, a $2d$ chiral boson is dual to $1/2$ level $3d$ $U(1)$ Chern Simons (CS) theory. To define this CS theory requires a quadratic refinement, which is naturally provided by spin structure. It is interesting to see if the modern techniques can also be generalized to   general $2d$ chiral boson systems, as well as potential bosonization in higher dimensions.  
\end{itemize}

\subsubsection*{Acknowledgments}
We thank Stefan Theisen for valuable discussions. The work of PC is supported by the Cluster of Excellence \textit{PRISMA}$^{++}$ (EXC 2118/2, Project ID 390831469). The work of HPF is supported in part by a grant from the Simons Foundation (602883,CV) and a gift from the DellaPietra Foundation.
\appendix

\section{Bordism group computations}
\label{app:bordism-computation}
In this appendix we present the computations of $\Omega_{3d}^{Spin,tor}(B\bZ_n\times B\bZ_2^F)$ used in the main text:
\begin{equation}
    \label{eq:bordism-groups}
    \begin{split}
        & \Omega_{3d}^{Spin,tor}(B\bZ_n\times B\bZ_2^F) = \Omega_{3d}^{Spin,tor}(B\bZ_n )\oplus \Omega_{3d}^{Spin,tor}(B\bZ_2^F)=\bZ_{n}\oplus \bZ_{8}, \ \ \text{$n$ odd}  \\&
        \Omega_{3d}^{Spin,tor}(B\bZ_n\times B\bZ_2^F) =  \Omega_{3d}^{Spin,tor}(B\bZ_n )\oplus  \Omega_{3d}^{Spin,tor}(B\bZ_2^F)\oplus K = \bZ_{2n}\oplus \bZ_{8}\oplus \bZ_2\oplus \bZ_2 \oplus \bZ_2, \ \ \text{$4|n$ }.
    \end{split}
\end{equation}
Equation \eqref{eq:bordism-groups} was already anticipated in \cite{Guo:2018vij}. That reference worked out the case $n=4$. Since for general $n$ divisible by $4$, $H^*(B\bZ_n,\bZ_2)$ has the same module structure over the mod-$2$ Steenrod subalgebra generated by $Sq^{1,2}$, the computation of $\Omega_{3d}^{Spin,tor}(B\bZ_n\times B\bZ_2^F)$ is similar to $n=4$ case. 

The main tools are the Atiyah--Hirzebruch spectral sequence (AHSS) and the Adams spectral sequence (ASS). Useful introductions to AHSS and ASS may be found in \cite{davis2001lecture,beaudry2018guidecomputingstablehomotopy}.
\subsection{ $\Omega_{3d}^{Spin,tor}(B\bZ_n\times B\bZ_2^F)$ with $n$ odd}
\label{subsec:bordism-odd-n}
To apply AHSS to this case, we first compute $H_{*}(B\bZ_{n}\times B\bZ_2;\bZ(\bZ_2))$. Denoting $X = B\bZ_n \times B\bZ_2$, first applying Kunneth formula to get:
\begin{equation}
    \begin{cases}
        H_0(X;\bZ)  =\bZ\\
        H_1(X;\bZ)  =\bZ_n\oplus \bZ_2\\
        H_2(X;\bZ)  =0\\
        H_3(X;\bZ)  =\bZ_n\oplus \bZ_2\\
        H_4(X;\bZ)  =0\\
        ...
    \end{cases}
\end{equation}
The universal coefficient theorem gives:
\begin{equation}
    \begin{cases}
        H_0(X;\bZ_2)  =\bZ_2\\
        H_1(X;\bZ_2)  = \bZ_2\\
        H_2(X;\bZ_2)  =\bZ_2\\
        H_3(X;\bZ_2)  =  \bZ_2\\
        H_4(X;\bZ_2)  = \bZ_2\\
        ...
    \end{cases}
\end{equation}
The second page of AHSS in this case gives
 \begin{equation}\label{eq:ahss-odd-n}
	\begin{array}{c}
	E^2_{p,q}=H_p\big(X;\Omega_q^{\text{spin}}\big)\\
	\begin{array}{c|cccccccc}
		6 &0&0&0&0&0 \\
		5 &0&0&0&0&0& \\
		4 & \bZ&  \bZ_n\oplus\bZ_2 &0&\bZ_n\oplus\bZ_2& 0 \\
		3 &0&0&0&0&0  \\
		2 & \bZ_2 &\bZ_2&\bZ_2& \bZ_2 &  \bZ_2 & \\
		1 & \bZ_2 &\bZ_2&\bZ_2&\bZ_2& \bZ_2 \\
		0 & \bZ&  \bZ_n\oplus\bZ_2 &0&\bZ_n\oplus\bZ_2& 0  \\
		\hline
		& 0 & 1 & 2 & 3 & 4  
	\end{array}
	\end{array}
\end{equation}
Focusing on the $p+q =3$ entries, we see the number of group elements $|\Omega_{3d}^{Spin,tor}(B\bZ_n\times B\bZ_2^F)|\leq 8n$. On the other hand, $X$ admits to fibrations
\begin{equation}
    \label{eq:two-fibrations}
    \begin{split}
        & B\bZ_n \xrightarrow{i_1} X \to B\bZ_2,\\&
         B\bZ_2 \xrightarrow{i_2} X \to B\bZ_n,
    \end{split}
    \end{equation}
and two projections
\begin{equation}
    \label{eq:two-projections}
    \begin{split}
        &\pi_1:X \longrightarrow B\bZ_n\\&
        \pi_2: X \longrightarrow B\bZ_2
        \end{split}
\end{equation}
which satisfies
\begin{equation}
\label{eq:identity-maps}
    B\bZ_{n }\xrightarrow{ \pi_{1}\circ i_{1} =id} B\bZ_n, \quad   B\bZ_{2}\xrightarrow{ \pi_{2}\circ i_{2} =id} B\bZ_2.
\end{equation}
As $\Omega_{*}^{Spin,tor}(*)$ is a generalized homology theory, apply its functoriality associated with $\pi_{i},i_{i},\pi_{i}\circ i_{i} =id$ and the fact $\Omega_{3d}^{Spin,tor}(B\bZ_n) = \bZ_n$ for $n$ odd, $\Omega_{3d}^{Spin,tor}( B\bZ_2^F) =\bZ_8$ gives:
\begin{equation}
    \label{eq:bordism-odd-n-prelim}
    \Omega_{3d}^{Spin,tor}(B\bZ_n\times B\bZ_2^F) = \Omega_{3d}^{Spin,tor}(B\bZ_n )\oplus \Omega_{3d}^{Spin,tor}( B\bZ_2^F)\oplus... = \bZ_8 \oplus \bZ_n\oplus...
\end{equation}
As $|\Omega_{3d}^{Spin,tor}(B\bZ_n\times B\bZ_2^F)|\leq 8n$, we arrive at:
\begin{equation}
    \label{eq:bordism-odd-n}
     \Omega_{3d}^{Spin,tor}(B\bZ_n\times B\bZ_2^F) = \Omega_{3d}^{Spin,tor}(B\bZ_n )\oplus \Omega_{3d}^{Spin,tor}(B\bZ_2^F)=\bZ_{n}\oplus \bZ_{8}, \ \ \text{$n$ odd} .
\end{equation}
\subsection{$\Omega_{3d}^{Spin,tor}(B\bZ_n\times B\bZ_2^F)$ with $4|n$}
\label{subsec:bordism-even-n}
First we compute $H_*(B\bZ_n\times B\bZ_2;\bZ (\bZ_2))$. Denote $Y = B\bZ_n \times B\bZ_2$, first applying Kunneth formula to get
\begin{equation}
    \begin{cases}
        H_0(Y;\bZ)  =\bZ\\
        H_1(Y;\bZ)  =\bZ_n\oplus \bZ_2\\
        H_2(Y;\bZ)  =\bZ_2\\
        H_3(Y;\bZ)  = \bZ_n\oplus \bZ_2 \oplus \bZ_2\\
        H_4(Y;\bZ)  = \bZ_2 \oplus \bZ_2\\
        ...
    \end{cases}
\end{equation}
The universal coefficient theorem gives
\begin{equation}
    \begin{cases}
        H_0(Y;\bZ_2)  =\bZ_2\\
        H_1(Y;\bZ_2)  = \bZ^{\oplus 2}_2\\
        H_2(Y;\bZ_2)  =\bZ^{\oplus 3}_2\\
        H_3(Y;\bZ_2)  =  \bZ^{\oplus 4}_2\\
            H_4(Y;\bZ_2)  = \bZ^{\oplus5}_2\\
        ...
    \end{cases}
\end{equation}
The second page of AHSS in this case gives
 \begin{equation}\label{eq:ahss-even-n}
	\begin{array}{c}
	E^2_{p,q}=H_p\big(X;\Omega_q^{\text{spin}}\big)\\
	\begin{array}{c|cccccccc}
		6 &0&0&0&0&0 \\
		5 &0&0&0&0&0& \\
		4 & \bZ&  \bZ_n\oplus\bZ_2 &\bZ_2&\bZ_n\oplus\bZ_2^{\oplus2}  & \bZ_2^{\oplus2}  \\
		3 &0&0&0&0&0  \\
		2 & \bZ_2 &\bZ_2^{\oplus2}&\bZ_2^{\oplus3}& \bZ_2^{\oplus4}& \bZ_2^{\oplus5} \\
		1 & \bZ_2 &\bZ_2^{\oplus2}&\bZ_2^{\oplus3}& \bZ_2^{\oplus4}& \bZ_2^{\oplus5}\\
		0 & \bZ&  \bZ_n\oplus\bZ_2 &\bZ_2&\bZ_n\oplus\bZ_2^{\oplus2}  & \bZ_2^{\oplus2}  \\
		\hline
		& 0 & 1 & 2 & 3 & 4  
	\end{array}
	\end{array}
\end{equation}
Again, using maps \eqref{eq:two-fibrations},\eqref{eq:two-projections}, \eqref{eq:identity-maps} and the fact $\Omega_{3d}^{Spin,tor}(B\bZ_n) =   \bZ_{2n}\oplus \bZ_2$, we have:
\begin{equation}
    \label{eq:bordism-even-n-prelim}
    \Omega_{3d}^{Spin,tor}(B\bZ_n\times B\bZ_2) = \bZ_{2n}\oplus \bZ_2 \oplus \bZ_8 \oplus K.
\end{equation}
In fact, using the fact that for $q=0,1$ the differential $d_2$ is dual to $Sq^2$, we see that all elements in $p+q =3$ entries at \eqref{eq:ahss-even-n} survive. Hence there are only two possibilities:
\begin{equation*}
    K = \bZ_4  \quad \text{or} \quad K=\bZ_2 \oplus \bZ_2.
\end{equation*}
Notice when $n=2$, $K=\bZ_4$. Next we use ASS to see that $K =\bZ_2 \oplus \bZ_2$ when $4|n$.

To use ASS, first we need to find the module structure from $H^*(B\bZ_{n}\times B\bZ_2)$ of Steenrod subalgebra generated by\footnote{This is enough as we only need spin bordism at dimension 3.} $Sq^1,Sq^2$. The corresponding module structures are:
\begin{equation}
    \label{eq:steenrod-module}
      \bZ_2 \oplus \Sigma^1 \bZ_2\oplus  \Sigma^1 S \oplus \Sigma^2 S \oplus \Sigma^2\cA_1/\cE_1\oplus \Sigma^3\cA_1/\cE_1 \oplus \Sigma^3 R_6 \oplus \Sigma^4 R_6.
\end{equation}
where the modules $\bZ_2,S,\cA_1/\cE_1,R_6$ and their Adams charts can be found in e.g. \cite{beaudry2018guidecomputingstablehomotopy}. 

\begin{figure}
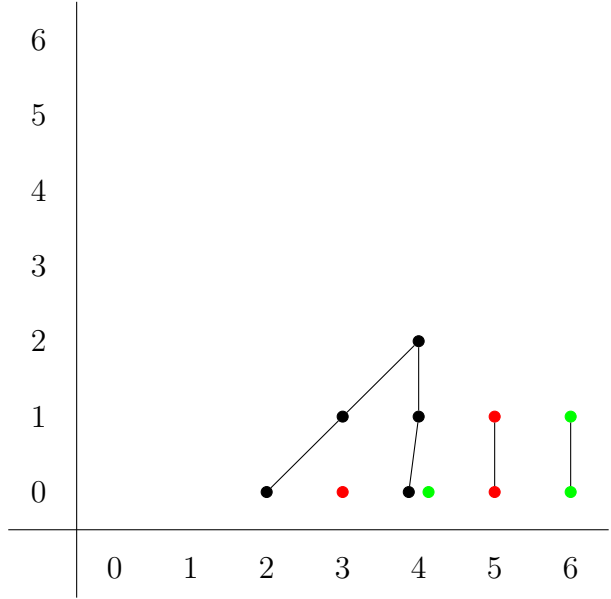

\begin{center}
\begin{sseqdata}[name=MMMM,Adams grading,classes = fill,xrange = {0}{6},yrange = {0}{6}]

	\class(2,0)
    \class(3,1)
    \class(4,2)
    \class(4,1)
    \class(4,0)
     \class[red](3,0)
      \class[red](5,0)
       \class[red](5,1)
       \class[green](4,0)
       \class[green](6,0)
       \class[green](6,1)
   \structline(2,0)(3,1)
    \structline(3,1)(4,2)
     \structline(4,2)(4,1)
      \structline(4,0)(4,1)
       \structline(5,0)(5,1)
         \structline(6,0)(6,1)
	 
\end{sseqdata}

\printpage[name = MMMM,page = 2]
\end{center}
\caption{Second page of ASS for $K \subset \Omega_{3d}^{Spin,tor}(B\bZ_n\times B\bZ_2^F)$ with $4|n$}
\label{fig:ass}
\end{figure}
Applying these adams charts to the module structure \eqref{eq:steenrod-module}, we read out the second page of ASS. Here we only draw the relevant pieces for computing the group $K$, thanks to \eqref{eq:bordism-even-n-prelim}, which are given by $\S^2 S\oplus\Sigma^3 R_6 \oplus \Sigma^4 R_6$. The second page of ASS is shown in Fig \ref{fig:ass}. It is clear that no differential will kill the red and black dots in the 3rd column and:
\begin{equation}
K=\bZ_2 \oplus \bZ_2.
\end{equation}

Also from Fig \ref{fig:ass} it is clear that the corresponding generators are given by the 3rd integral group homology $H_3(B\bZ_n\times B\bZ_2;\bZ)$ (the red dot) and a direct product of a circle times a two dimensional manifold with proper $\bZ_n \times \bZ_2$ bundle structure.

\bibliographystyle{JHEP}
\bibliography{anomalies}
\end{document}